# Comparative analysis of neural network architectures for short-term FOREX forecasting

**THEODOROS ZAFEIRIOU**

0000-0001-7277-8768

*Hellenic Open University, Parodos Aristotelous 18*
*Patras, 26335, Greece*
*zafiriou.theodore@ac.eap.gr*

**DIMITRIS KALLES**

0000-0003-0364-5966

*Hellenic Open University, Parodos Aristotelous 18*
*Patras, 26335, Greece*
*kalles@eap.gr*

**Abstract**

The present document delineates the analysis, design, implementation, and benchmarking of various neural network architectures within a short-term frequency prediction system for the foreign exchange market (FOREX).

Our aim is to simulate the judgment of the human expert (technical analyst) using a system that responds promptly to changes in market conditions, thus enabling the optimization of short-term trading strategies.

We designed and implemented a series of LSTM neural network architectures which are taken as input the exchange rate values and generate the short-term market trend forecasting signal and an ANN custom architecture based on technical analysis indicator simulators

We performed a comparative analysis of the results and came to useful conclusions regarding the suitability of each architecture and the cost in terms of time and computational power to implement them. The ANN custom architecture produces better prediction quality with higher sensitivity using fewer resources and spending less time than LSTM architectures. The ANN custom architecture appears to be ideal for use in low-power computing systems and for use cases that need fast decisions with the least possible computational cost.

## 1. Introduction

The majority of profits in the foreign exchange market, particularly in FOREX [01], are derived from extensive leverage utilizing margin [02]. Leverages reaching as high as 1:200 (meaning someone with an initial capital of €1000 can risk capital of €200,000) pose a significant risk for low-volatility investments, particularly those conducted on the same day and sometimes within a few minutes. Consequently, there is a contention that forecasting models [03] should be grounded in short time periods.

In markets characterized by substantial depth and volume, such as FOREX, capitalizing on micro-volatility in the short term holds paramount importance and can be accomplished through analogous short-term forecasts [04].

Over the past few decades, economists have endeavored to construct models capable of successfully predicting trends, giving rise to the field known as technical analysis. Despite extensive and prolonged efforts, there is still no universally applicable index or model that can reliably forecast financial market trends. The primary obstacle stems from technical analysis neglecting the most recent shifts in fundamentals, which remain unrecorded, as well as the impact of breaking news on investor psychology.

The aim of this paper is to compare the short-term trend prediction provided by a number of artificial neural network architectures by drawing useful conclusions about the suitability of each of them.

Specifically, we compare the quality of the prediction of six different parameterizations of vanilla LSTM, bidirectional LSTM and convolutional LSTM networks with a prototype artificial neural network architecture based on simple error backpropagation networks.

This paper is structured in four sections. First, we briefly review related work on exchange rate forecasting using computational intelligence. We then describe the architectures of the different forecasting systems in our experimentation and proceed to present and analyze the experimental results before concluding, in the last section, where we also set out some future directions for work.

## 2. A brief background on predicting exchange rates using computational intelligence

As previously mentioned, traders in the FOREX market utilize technical analysis tools [05] to forecast exchange rates. However, automated systems [03] often yield higher profits by trading substantial sums based on forecasting models. The success of technical analysis methods varies, with failures typically attributed to undetected changes in fundamental values and market psychology. Forecasting inaccuracies tend to increase with shorter-term forecasting [06].

Efficiently approaching the challenge of automated trading with a large portfolio strategy that continuously processes data streams across diverse markets is demonstrated in [07]. The paper introduces a scalable trading model that learns to generate profit from multiple inter-market price predictions and market correlation structures.

Forecasting methods are broadly categorized into traditional and non-traditional approaches. Traditional methods rely on static algorithms unaffected by input data [08], serving as econometric models for result interpretation and hypothesis testing, a standard quality assurance procedure in technical analysis [08].

Non-traditional methods, on the other hand, encompass data-driven approaches that self-correct [08]. These methods, such as fuzzy logic [09], Artificial Neural Networks (ANN) [10], neuro-fuzzy architecture (hybrid systems) [11], and genetic algorithms [12], can be competitive with econometric methods due to their generalized operations [13]. Machine-learning-based methods, particularly those using past trading data, are considered robust for predicting trading patterns in FOREX [10].

Neural networks, especially those with hidden layers, offer an internal representation of variable relationships and excel in handling sparse data and complex phenomena [14]. Genetic algorithms have been employed to learn trading rules and combined with echo-state networks for market trend prediction, yielding better results in both bull and bear markets compared to conventional strategies [15].

We now briefly review some key contributions to the field.

Cavalcante et al.'s [16] comprehensive overview of primary studies from 2009 to 2015, emphasizing techniques for preprocessing, clustering financial data, forecasting market movements, and mining financial information. Patel et al. [17] focused on predicting stock market index prices using a two-stage fusion approach with Support Vector Regression (SVR) and Artificial Neural Networks (ANN). Yıldırım et al.

[18] utilized LSTM networks for directional predictions in Forex, achieving success with a hybrid model incorporating macroeconomic and technical indicator data.

Fisher et al. [19] employed LSTM networks for predicting directional movements in S&P 500 constituent stocks, with varying profitability over time. Xiong et al. [20] applied Long Short-Term Memory neural networks to model S&P 500 volatility, outperforming linear benchmarks. Galeshchuk and Mukherjee [21] investigated the use of deep convolutional neural networks for predicting exchange rate direction with satisfactory accuracy.

In previous work [22], we developed an ANN to predict market signals in the FOREX, combining advantages of technical analysis and ANN in causal modeling and case control. In a subsequent study [23], we presented an ultra-short-term frequency trading system for FOREX, incorporating artificial intelligence techniques for pre-trade analysis, trend forecasting, and trade execution. The system aimed to simulate human expert judgment and decision-making, achieving superior performance compared to individual or combined technical indicators across various automated trading engines.

In this paper we experiment with several LSTM network architectures and compare their performance with the performance of an improved version of the aforementioned architecture, drawing useful conclusions about their suitability for FOREX time series prediction.

## 3. A detailed system description

In this section, we detail our approach to analyzing, designing, and implementing ultra-short trend prediction. This system encompasses crucial stages, namely Pretrade Analysis and Transaction Signal Production (Trend Forecasting) [24].

Our objective is to emulate the decision-making of a human expert, whether a technical analyst or broker, through an artificial intelligence system that adeptly responds to changes in market conditions. This responsiveness is integral to optimizing the efficiency of short-term transactions.

The analysis stage commences with data mining, where relevant data for subsequent steps are carefully selected. Subsequently, in the trend forecasting stage, various Artificial Neural Network (ANN) architectures are conceived and implemented to generate trend forecasting signals. The final step involves a comparative analysis of different sources of trend forecasting, specifically the diverse ANN architectures employed.

### 3.1. *Selection of the Exchange rate and experimental data source*

For our experiments, we opted to focus on the EUR/USD exchange rate, given its status as the world's largest trading currency pair. The market depth of this pair acts as a deterrent to lobbies engaging in price manipulations that could distort its true representation.

Our selected sources for experimental data include Truefx [25], recognized as an industry-leading exchange rate data server, and American Integral [26]. Integral is utilized by the largest institutional service FOREX providers globally for their price references.

The experimental data pertains to the tick-to-tick EUR/USD exchange rate for the months of October, November, and December 2021. Initially, the dataset comprises over 10 million values, which undergo pre-processing to eliminate flat areas where the exchange rate remains constant.

### 3.2. *Selected LSTM networks*

Recurrent networks leverage feedback connections to retain information from recent input events as a trigger for the activation function, enabling the incorporation of short-term memory. Although networks of this type are effective for several applications (e.g. voice recognition) they have weaknesses in cases where there is a non-trivial time lag between the input and the expected output.

"Long Short-Term Memory" or LSTM networks, commonly known as such, are recurrent networks specifically designed to address the issue of rapidly diminishing short-term memory in retaining information over longer sequences. The LSTM model effectively preserves selected information in long-term memory, which is stored in the cell state, while short-term information is captured in the hidden state.

For the implementation of the chosen LSTM architectures, we utilized Keras & TensorFlow 2. Keras is a deep learning API written in Python, operating on the TensorFlow machine learning platform. It is designed with a focus on facilitating rapid experimentation [27]. Known for its top-notch performance and scalability.

We selected eight different LSTM architectures for our experimentation with parameters as shown in Table 1. All these LSTM architectures follow the sequential model and have a ReLU activation function.

**Table 1**. Selected LSTM architectures

| Name | LSTM Units | Dense Units | Lookback * | Bidirectional | Convolutional |
|---|---|---|---|---|---|
| sLSTM-1-1 | 100 | 1 X 1 | 1 | No | No |
| sLSTM-15-1 | 100 | 1 X 1 | 15 | No | No |
| sLSTM-15-1,15 | 100 | 1 X 15, 1 X 1 | 1 | No | No |
| biLSTM-1-1 | 100 | 1 X 1 | 1 | Yes | No |
| biLSTM-15-1 | 100 | 1 X 1 | 15 | Yes | No |
| biLSTM-15-1,15 | 100 | 1 X 15, 1 X 1 | 15 | Yes | No |
| convLSTM-1-1 | 60 | 1 X 1 | 1 | No | Yes |
| convLSTM-1-1,15 | 64 | 1 X 1 | 15 | No | Yes |

* The number of sequences of input LSTM will train before generating an output.

In the following figures we show the various LSTM architectures.

**Figure 1**. sLSTM-1-1 and sLSTM-15- architectures

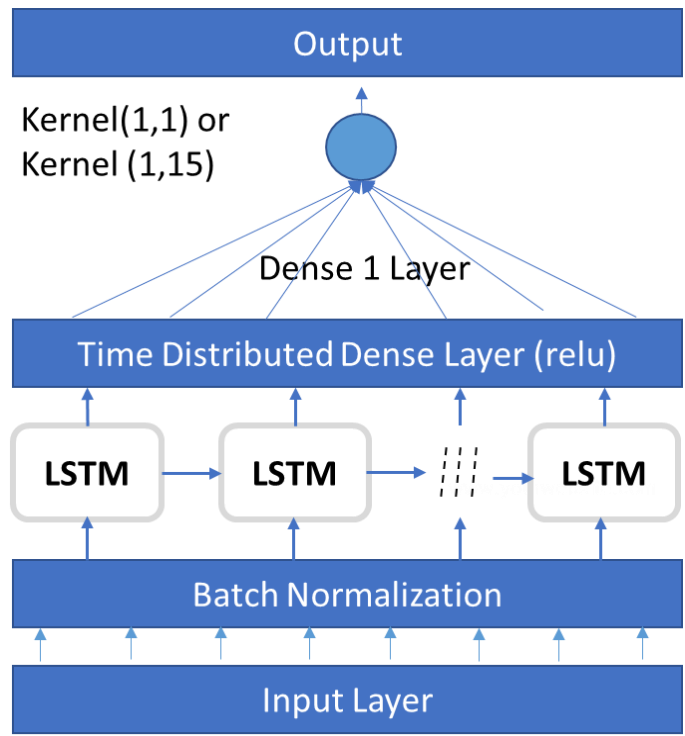

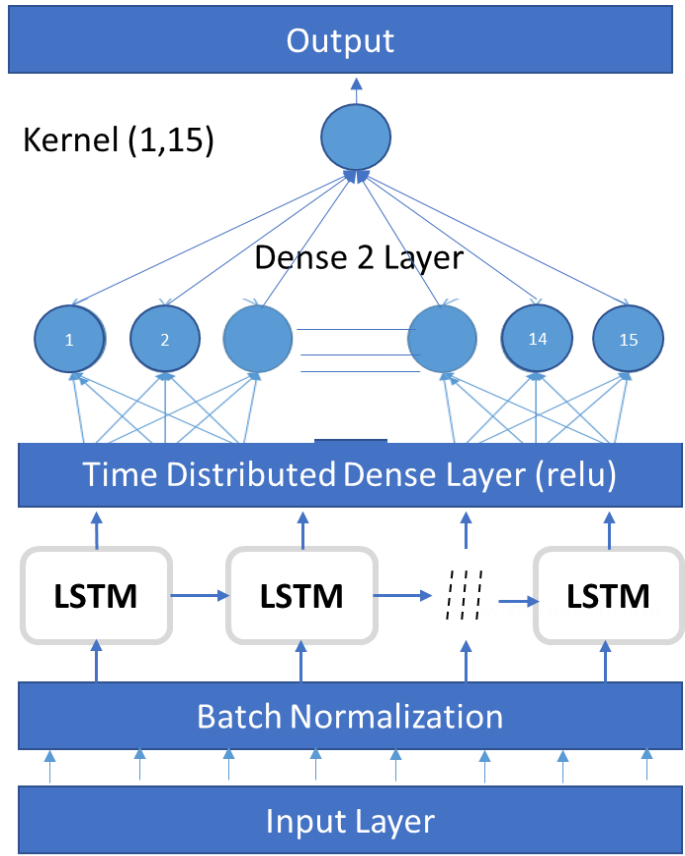

**Figure 2**. sLSTM-15-1 and sLSTM-15-1,15 architectures.

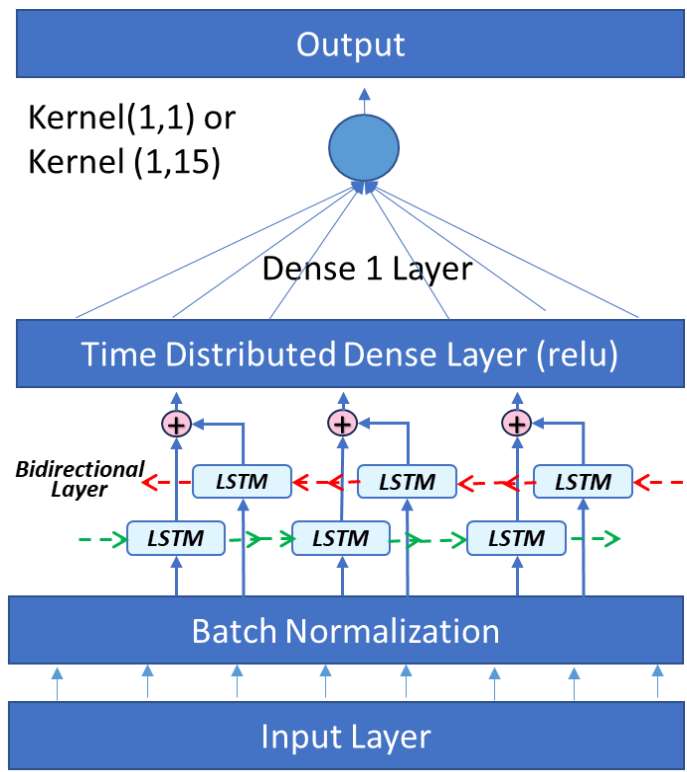

**Figure 3.** biLSTM-1-1 and biLSTM-15- architectures

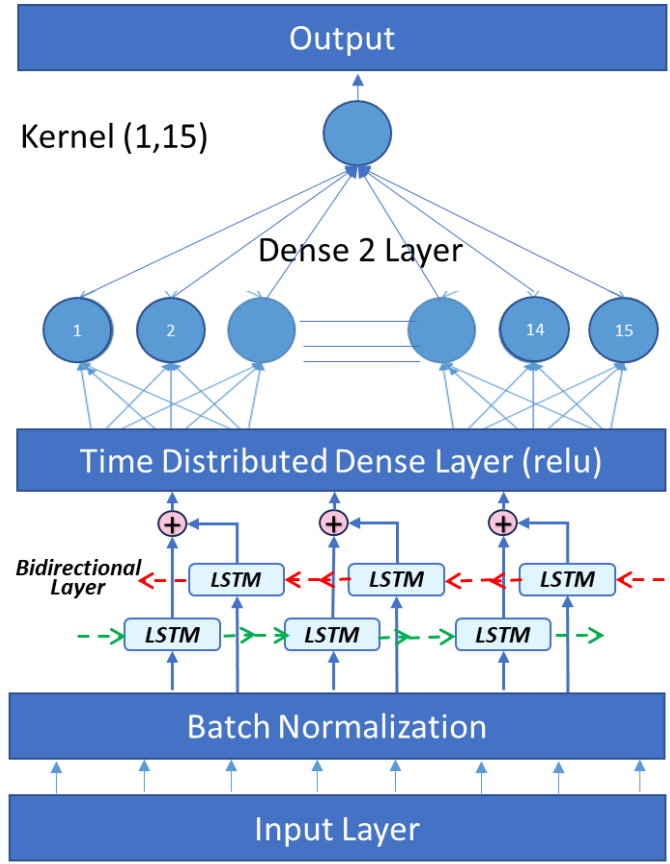

**Figure 4.** biLSTM-15-1 architecture

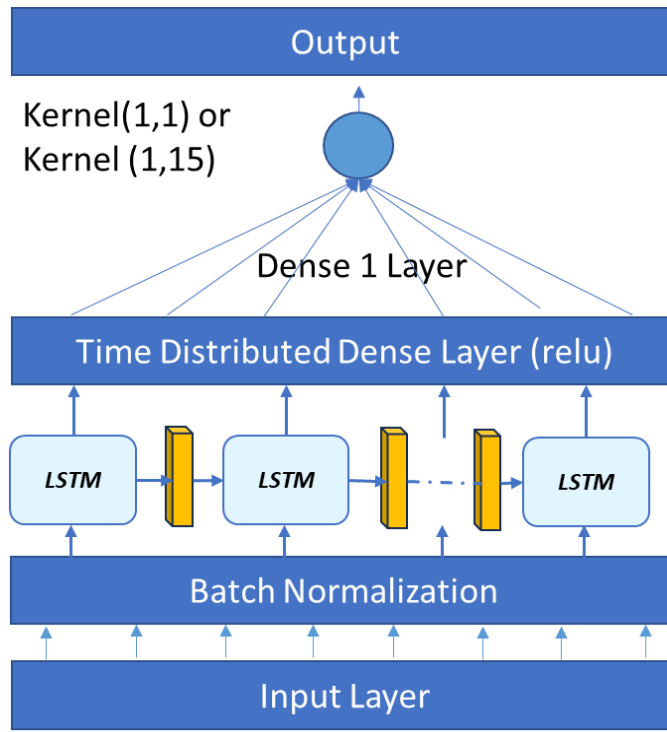

**Figure 5.** convLSTM-1-1 and convLSTM-1-1,15 architectures.

## 3.3. *ANN architecture based on technical analysis indicator simulators*

After initial experimentation, we meticulously chose and adapted specific algorithms for our technical indicators in the experiments [23], aligning with short-term forecasting objectives (Figure 6). These include modified arithmetic moving averages (MAs) calculated over 300, 600, and 900 price intervals, the RSI-300 oscillator, the CCI-300 oscillator, the Williams-300 oscillator, and the Price Oscillator (MA-300, MA-600, MA-900). The application of these technical indicators generates forecasts outlined in Annex I.

The input parameters encompass exchange rates, time, and dates (Figure 6). The system, utilizing the predicted trend signal and its auto-trading agents' configurations, engages in simulated short-term trading, generating performance logs that simulate profit or loss.

Data inputs are utilized in the custom technical indicator simulators (Figure 6) [28]. Each technical indicator simulator yields an output from the set of values detailed in Table 2. The outputs from these simulators are directed to the input neurons of the ANN system, as extensively documented in previous studies [23].

**Figure. 6. An overview of the system – architecture**

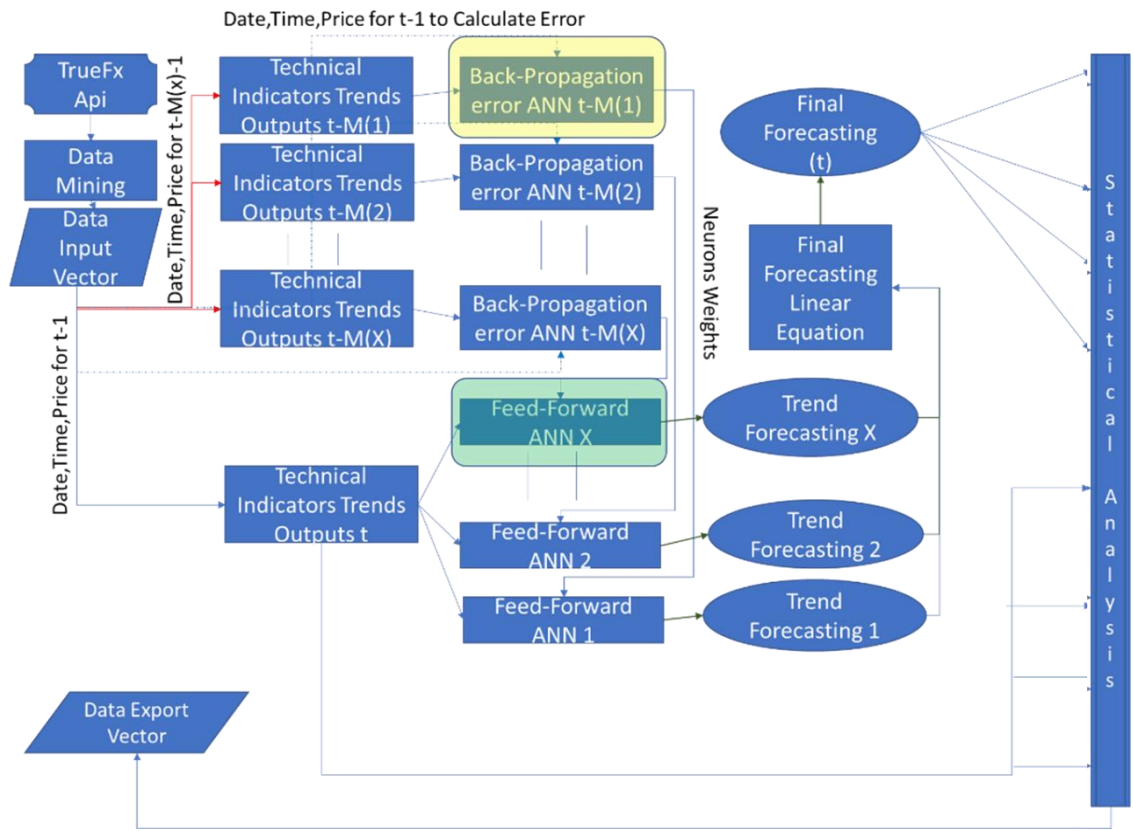

The prediction system comprises two sets of Artificial Neural Networks (ANNs) operating in pairs. In each pair, one ANN receives the outputs of simulators corresponding to technical indicators as inputs and operates in conventional error back-propagation mode, striving to align with the trend prediction. This ANN, utilizing past values, calculates the prediction error. The learned weights from this ANN are then transferred to its paired ANN. However, the paired ANN operates exclusively in feed-forward mode, considering present values. Thus, one ANN is trained on historical data, while its counterpart generates predictions on current data. All feed-forward ANNs are combined in an ensemble to generate the final trend forecast [23]. This architecture is a modification of the fundamental Generative Adversarial Network concept [29].

Custom technical indicators are created, and their predicted trends at time t-M(x)-1 are sent to the input layer of each back-propagation ANN. Each technical indicator corresponds to an input neuron of the ANN, with its calculation reflecting its value at time t-M(x)-1. Here, t-M(x) signifies the time at which the neural network with index x operated in the past (e.g., M(1) = 30 indicates a focus on confirming the technical indicator's prediction within 30 seconds). The hidden layer employs a tanh-type sigmoid activation to produce output values in the range of [-2, +2], while the output layer is linear. The number of hidden layer neurons is set at double the number of input layer neurons based on preliminary results. The output layer neurons, along with corresponding data, export the trend signal for each back-propagation ANN (Figure 6) [23].

Furthermore, the algorithm for calculating the real trend is updated using data from time points t-1 and t-1-M(x) (Table 3). This algorithm generates a normalized estimated value of the real trend. The output value of the final node is then compared to the real trend to train the neural network. Real trend conditions (Table 3) are selected after preliminary experimentation.

Each back-propagation ANN in the series is characterized by the time it operates in the past (t-M(x)), with the number of back-propagation ANNs being configurable. The number of feed-forward ANNs equals the number of back-propagation ANNs, as each back-propagation ANN feeds the weights of its neurons into a corresponding feed-forward ANN [23].

Custom technical indicators are generated, and their predicted trends for time t are sent to the input layer of each feed-forward ANN. The hidden layer employs a tanh-type sigmoid activation to produce output values in the range of [-2, +2], while the output layer is linear. All neuron weights are fed from the neuron weights of a corresponding back-propagation ANN [23].

Each back-propagation ANN in the series is characterized by the time it operates in the past (t-M(x)), with the number of back-propagation ANNs being customizable. The count of feed-forward ANNs matches the number of back-propagation ANNs, as each back-propagation ANN channels the weights of its neurons into a corresponding feed-forward ANN [23].

Custom technical indicators are generated, and their predicted trends for time t are transmitted to the input layer of each feed-forward ANN. The hidden layer, activated by a tanh-type sigmoid, produces output values within the range of [-2, +2], while the output layer is linear. All neuron weights are derived from the neuron weights of a corresponding back-propagation ANN [23].

**Table 2. Mapping of numerical values to trends.**

| Value | Corresponding trend |
| --- | --- |
| +2 (-2) | Absolutely positive (Absolutely negative) |
| +1.5 (-1.5) | Quite positive (Quite negative) |
| +1 (-1) | Positive (Negative) |
| +0.5 (-0.5) | Neutral / positive (Neutral / negative) |
| 0 | Neutral |

Each Forecasting Trend (FT(x)) from the ANN feedforward series contributes a certain proportion to the final Forecasting Trend (FFT) of the system (Figure 7). This algorithm essentially determines the contribution weight of each ANN feedforward to the ultimate forecast. The contribution of each ANN to the final prediction is calculated as the inverse of its absolute error divided by the sum of the inverses of the absolute errors of both ANNs for time t-K, where K=0,1,2,3,... (before training the ANNs for time t-K). The FFT is then normalized to one of the values shown in Table 2 [23].

The parameter values for all neural networks (both back-propagation and feed-forward series) were chosen based on our previous work to ensure comparability (Table 4). Similar to our prior work, each series of back-propagation and feed-forward ANNs consists of three ANNs (three pairs of ANNs). Additionally, each back-propagation ANN has five parameters, as outlined in Table 5. The parameter values for the technical analysis simulators align with those in our previous work (Table 6). The Predicted Trend Value defines the upward or downward multiplier of the exchange rate required for the neural network to trigger the corresponding trend (±2, ±1.5, ±1, ±1, ±1, ±0.5). Essentially, the rate of rise or fall of the exchange rate characterizes the market trend as neutral, slightly bullish/bearish, bullish/bearish, quite bullish/bearish, very bullish/bearish. The predicted trend values (Table 5) are chosen after preliminary experimentation [23].

**Figure 7. An overview of the calculation of the final forecasting trend (for two networks).**

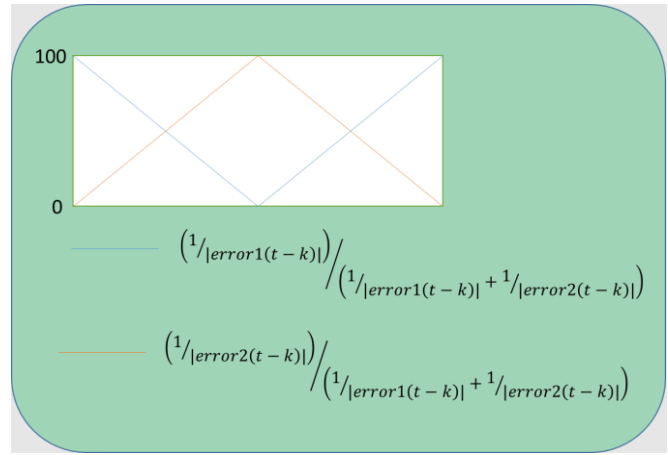

**Table 3. Conditions for the actual trend in the forecasting trend signal of the current ANN system (rules are listed in descending order of priority) are in line with ultra-short-term trading.**

| Conditions | Value |
| --- | --- |
| price(t-1)/price(t-1-M(x))>1,00090 | +2 |
| price(t-1-M(x))/price(t-1)>1,00090 | -2 |
| price(t-1)/price(t-1-M(x))>1,00060 | +1,5 |
| price(t-1-M(x))/price(t-1)>1,00060 | -1,5 |
| price(t-1)/price(t-1-M(x))>1,00030 | +1 |
| price(t-1-M(x))/price(t-1)>1,00030 | -1 |
| price(t-1)/price(t--1M(x))>1,00015 | +0,5 |
| price(t-1-M(x))/price(t-1)>1,00015 | -0,5 |
| Other Cases | 0 |

**Table 4. Parameterization of the Artificial Neural Network (ANN)**

| A/A | Parameter | Value |
|---|---|---|
| 1 | Number of ANN Epochs | 10 |
| 2 | Number of ANN Hidden Neurons | 14 |
| 3 | Learning rate of synapses between Hidden layer Neurons and Input Layer Neurons (LR-Inputs) | 0.001 |
| 4 | Learning rate of synapses between Hidden layer Neurons and Output Layer Neurons (LR-Output) | 0.001 |
| 5 | Number of Hidden Layers | 1 |
| 6 | Number of Output Neurons | 1 |
| 7 | Number of Input Neurons | 7 |
| 8 | Period (in a number of values) of auxiliary MA | 10 |

**Table 5. Parameterization of back-propagation ANN's.**

| A/A | Parameter Description | Values | | |
| | | ANN-1 | ANN-2 | ANN-3 |
|---|---|---|---|---|
| 1 | M(x) (In a number of prices approx. 1price =1sec) | 30 | 60 | 90 |
| 2 | Predicted Trend Value ±2 | 1,0090 | 1,0090 | 1,0090 |
| 3 | Predicted Trend Value ±1,5 | 1,0060 | 1,0060 | 1,0060 |
| 4 | Predicted Trend Value ±1 | 1,0030 | 1,0030 | 1,0030 |
| 5 | Predicted Trend Value ±0,5 | 1,0015 | 1,0015 | 1,0015 |

**Table 6. Parameterization of Technical Indicators Simulators.**

| A/A | Parameter Description | Value |
|---|---|---|
| 1 | Period (in a number of values) Oscillator RSI | 300 |
| 2 | Period (in a number of values) Oscillator Williams | 300 |
| 3 | Period (in a number of values) Oscillator CCI | 300 |
| 4 | Period (in a number of values) Short-Term MA | 300 |
| 5 | Period (in a number of values) Mid-Term MA | 600 |
| 6 | Period (in a number of values) Long-Term MA | 900 |
| 7 | Auxiliary Moving Averages of Price Oscillator | (300, 600, 900) |

### 3.4. *System development and use*

This ANN was developed in Java using the *Apache NetBeans* IDE 13.0 [30]. The application is fully configurable via a properly labeled parameter file. The LSTM architectures were developed in Python using *Google Colab* [31].

### 3.5. *Experimentation and results*

In this section, we will evaluate and compare various LSTM architectures outlined in Table 1 with the specific ANN architecture we have developed (Section 3.3). We will compare them in terms of forecasting success in the same field of experimentation, sensitivity in terms of the ability to generate forecasting signals and in terms of resource consumption.

Our experimentation involves the tick-to-tick EUR/USD exchange rate data for the months of October, November, and December 2021.

We used 2 metrics to compare the different architectures in our experimentation: Success in terms of trend of all prediction signals (STA) and success in terms of trend of strong prediction signals (STS) only. A strong prediction signal is considered to be a signal with an intensity <=-1 or >=1, as described in Table 3. A signal is considered successful in terms of direction and strength when it is confirmed within 900 exchange rate values (approximately 15 minutes). The conditions of success for each signal is shown in Table 7.

**Table 7. Condition of Success in term of trend of each value of signal.**

| Conditions for Success | Value of Signal |
|---|---|
| $\exists \left( price(t)/price(t+ (X \in (1,899))) \right) > 1,00090$ | +2 |
| $\exists \left( price(t+ (X \in (1,899)))/price(t) \right) > 1,00090$ | -2 |
| $\exists \left( price(t)/price(t- (X \in (1,899))) \right) > 1,00060$ | +1,5 |
| $\exists \left( price(t+ (X \in (1,899)))/price(t) \right) > 1,00060$ | -1,5 |
| $\exists \left( price(t)/price(t+(X \in (1,899))) \right) > 1,00030$ | +1 |
| $\exists \left( price(t+(X \in (1,899))/price(t) \right) > 1,00030$ | -1 |
| $\exists \left( price(t)/price(t+(X \in (1,899))) \right) > 1,00015$ | +0,5 |
| $\exists \left( price(t+(X \in (1,899))/price(t) \right) > 1,00015$ | -0,5 |

The data were fed to eight (8) LSTM architectures (Table 1) and our architecture described in Section 3.3. For the LSTM architectures, 50% of each month's data was used for training and 50% for trend forecasting. In our architecture, which is retrained serially with each new value, no training dataset was larger than the period of the long-term technical indicator used (in this case 900 exchange rate values, about 15min of data) is required. To make the results of our architecture and the LSTM architectures comparable, we present the trend forecast only for the data that predicted the trend and the LSTM architectures (2nd half of each month).

Table 8 shows the aggregate results of the experiment for the different LSTM architectures (section 3.2) and our ANN architecture.

**Table 8. Aggregated results of experimentation.**

|  | OCTOBER | | NOVEMBER | | DECEMBER | |
|---|---|---|---|---|---|---|
| **ANN** | STA | STS | STA | STS | STA | STS |
| Successful Forecasting Signals | 3808 | 310 | 10923 | 880 | 10989 | 437 |
| Total forecasting signals | 4641 | 407 | 13371 | 1070 | 13689 | 593 |
| % Success | **82,05%** | **76,17%** | **81,69%** | **82,24%** | **80,28%** | **73,69%** |
| **sLSTM-1-1** | | | | | | |
| Successful Forecasting Signals | 761 | 101 | 831 | 161 | 1419 | 253 |
| Total forecasting signals | 1091 | 161 | 1133 | 237 | 1921 | 424 |
| % Success | **69,75%** | **62,73%** | **73,35%** | **67,93%** | **73,87%** | **59,67%** |
| **sLSTM-15-1** | | | | | | |
| Successful Forecasting Signals | 769 | 96 | 483 | 80 | 1334 | 224 |
| Total forecasting signals | 1122 | 158 | 653 | 115 | 1803 | 372 |
| % Success | **68,54%** | **60,76%** | **73,97%** | **69,57%** | **73,99%** | **60,22%** |
| **sLSTM-15-1,15** | | | | | | |
| Successful Forecasting Signals | 782 | 100 | 310 | 58 | 1393 | 248 |
| Total forecasting signals | 1133 | 164 | 416 | 80 | 1892 | 418 |
| % Success | **69,02%** | **60,98%** | **74,52%** | **72,50%** | **73,63%** | **59,33%** |
| **biLSTM-1-1** | | | | | | |
| Successful Forecasting Signals | 779 | 105 | 760 | 142 | 1413 | 249 |
| Total forecasting signals | 1122 | 167 | 1033 | 213 | 1915 | 420 |
| % Success | **69,43%** | **62,87%** | **73,57%** | **66,67%** | **73,79%** | **59,29%** |
| **biLSTM-15-1** | | | | | | |
| Successful Forecasting Signals | 848 | 113 | 462 | 77 | 1344 | 238 |
| Total forecasting signals | 1244 | 197 | 621 | 109 | 1823 | 401 |
| % Success | **68,17%** | **57,36%** | **74,40%** | **70,64%** | **73,72%** | **59,35%** |
| **biLSTM-15-1,15** | | | | | | |
| Successful Forecasting Signals | 821 | 110 | 289 | 50 | 1397 | 259 |
| Total forecasting signals | 1199 | 191 | 378 | 68 | 1909 | 439 |
| % Success | **68,47%** | **57,59%** | **76,46%** | **73,53%** | **73,18%** | **59,00%** |
| **convLSTM-1-1** | | | | | | |
| Successful Forecasting Signals | 781 | 107 | 968 | 203 | 1350 | 240 |
| Total forecasting signals | 1125 | 169 | 1330 | 314 | 1829 | 402 |
| % Success | **69,42%** | **63,31%** | **72,78%** | **64,65%** | **73,81%** | **59,70%** |
| **convLSTM-1-1,15** | | | | | | |
| Successful Forecasting Signals | 352 | 37 | 106 | 24 | 894 | 104 |
| Total forecasting signals | 471 | 51 | 148 | 36 | 1179 | 165 |
| % Success | **74,73%** | **72,55%** | **71,62%** | **66,67%** | **75,83%** | **63,03%** |

We see that, in all three months, in both the STA and STS indices, the ANN custom architecture outperforms the LSTM architectures in terms of success rates. Furthermore, we observe that the absolute number of forecasting signals yielded by the specific ANN architecture is always more than 100% larger than the number of signals yielded by the LSTM architectures, suggesting a significantly higher sensitivity and better forecasting ability.

Figures 8 and 9 show the cumulative time series of successful STA and STS predictions over the entire experimentation.

Throughout the experiment it is clearly shown that the ANN-specific architecture outperforms all alternative LSTM architectures. There is no time window during which the predictions of the ANN-specific architecture produces inferior quality forecasting compared to the forecasting of the LSTM architectures. Note that the superiority of this ANN architecture is even more significant when we consider that it generates multiple numbers of forecasting signals than LSTM architectures.

The clarity of the picture is obvious as time passes.

At the end of the field experimentation the specific ANN architecture has produced 31,701 forecasting signals with 25,720 (81.13%) of them being successful. Correspondingly it has produced 2,070 strong forecasting signals of which 1,627 (78.6% percentage) were confirmed.

All LSTM architectures had similar performance between them. We can say that the basic LSTM architecture sLSTM-1-1 (Table 1) had the best relative performance by producing 4,145 forecasting signals of which 3,011 (72.64%) were successful. Correspondingly it produced 822 strong forecasting signals of which 515 (62.7%) were confirmed.

Therefore, the specific ANN architecture produced a total of 7.6 times more forecasting signals than LSTM. The successful forecasting signals of the specific ANN architecture are 8.5 times more than the LSTM architecture.

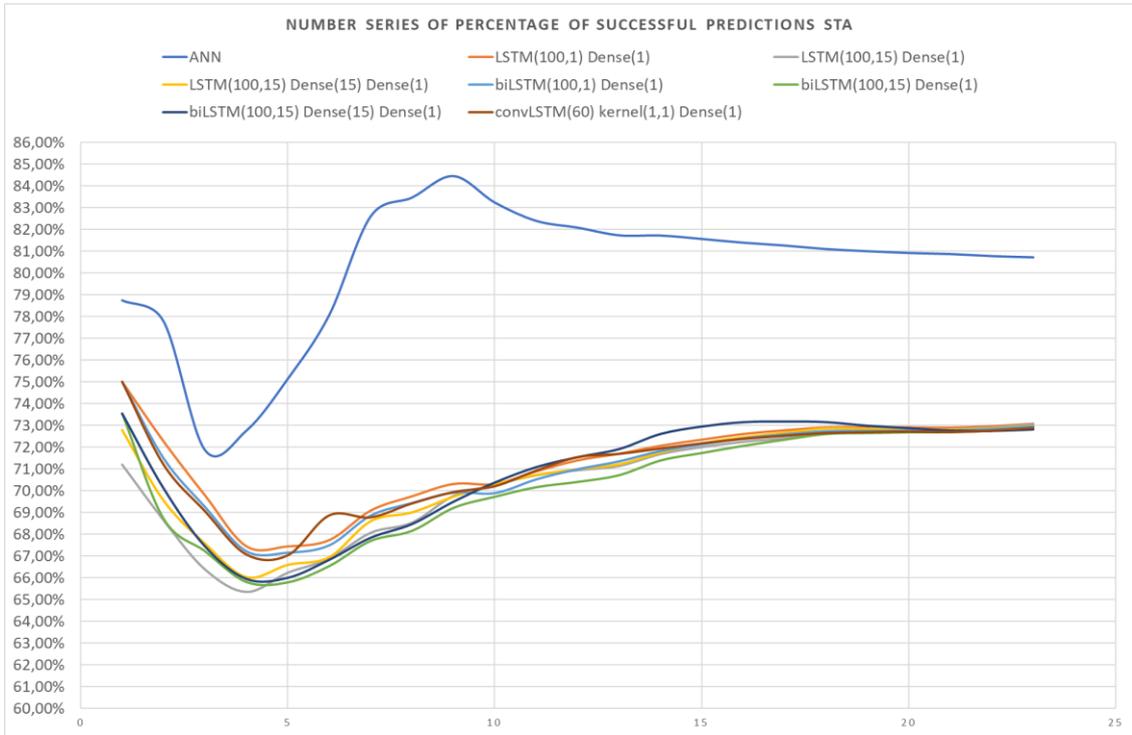

**Figure. 8.** Cumulative time series of the percentage of successful predictions - STA

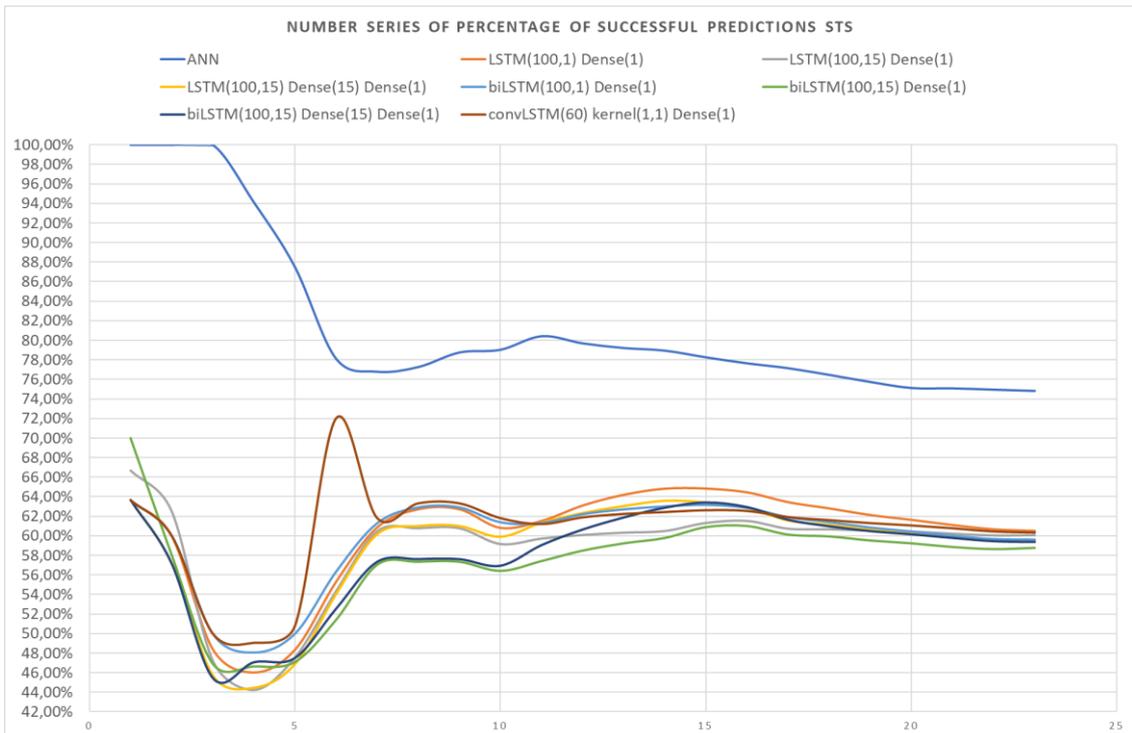

**Figure 9.** Cumulative time series of the percentage of successful predictions - STS

On a separate but increasingly important aspect, we note that when selecting an artificial neural network architecture one cannot fail to consider the resources it consumes to train and produce prediction.

For the medium complexity LSTM architecture of our experiments (biLSTM-15-1.15) using resources from google colab and specifically a python 3 Google Compute Engine backend with GPU acceleration, the time taken to train and predict the month of December 2021 was 1,175 seconds.

The specific ANN architecture using local resources (laptop with Ryzen 5 7520U processor without GPU acceleration), the time taken for same field of experimentation was 44 sec.

Therefore, the specific ANN architecture needs 27 times less time and fewer resources to perform the same field of experimentation compared to the LSTM architectures which can be an important selection criterion for users who cannot invest in the processing and communication overheads required by some modern cloud services.

### 3.6. *Conclusions*

We have designed and built a ANN custom architecture which combines machine learning and technical analysis.

Specifically, a set of modified artificial indicators are fed to the input neurons of an ANN architecture, which consists of a series of back propagation trained ANNs and a series of feed forward only ANNs, all of which work in pairs. In each pair, a backpropagation neural network (*learn-only* network) assigns its weights to an artificial neural network (*use-only* network) which only works in feedforward mode. The final prediction is based on a weighting algorithm that takes into account the prediction quality of each pair of neural networks (learn-only NN and use-only NN) of the previous time window.

The prediction quality of the custom architecture was compared with the prediction quality of 9 different LSTM architectures. We looked at both the absolute number of successful forecasts and the success rate of all of them. In all cases the custom ANN architecture outperformed the LSTM ones, producing a total of 31,701 forecasting signals with 25,720 (81.13%) of them being successful. The best performing LSTM architectures produced a total of 4,145 forecasting signals with 3,011 (72.64%) of them being successful. It becomes clear that the custom ANN architecture produces better quality forecasting while also being more sensitive, i.e. it produces more, better quality, trend signals

It is also important to note that our custom architecture trains and generates signals serially throughout the experimentation, requiring minimal initial calibration data depending on the maximum period of the modified technical indicators. Note that all LSTM architectures require training on the initial 50% of the experimentation field in order to then generate a reasonable forecast for the remaining 50%.

An increasingly important issue in the selection of an artificial neural network architecture is the resources it consumes to train and produce prediction. We have produced an indicative estimate that the custom ANN architecture requires nearly $1/25^{th}$ of the time and far fewer resources to perform the same field of experimentation compared to the LSTM architectures. This makes it possible to use it in real time devices with low computational resources, thus lowering the entry threshold for stakeholders who might want to join the FOREX trading market, as well as for other types of applications which rely on nearly-real-time data processing.

**Conflict of Interest**

The authors declare that they have no conflict of interest.

**Data Availability Statement**

The data that support the findings of this study are available from the corresponding author, upon reasonable request.

**Acknowledgement**

Some of the tables presented in this work are copied from work previously published by the authors so as to render the current paper self-contained. Proper citations and references have been included to attribute credit to the source of these tables, fully acknowledging the authors' earlier contributions to the field.

# ANNEX 1

Algorithms to calculate the Trend Forecasting of technical Indicators.

The tables are read from top to bottom. The first condition that is verified is valid

**Simulators of Moving Averages**

| Conditions | Trend Forecasting Signal |
|---|---|
| MA_M(t) < MA_10(t) && MA_M(t-1) ≥ MA_10(t-1) | +2 |
| MA_M(t) > MA_10(t) && MA_M(t-1) ≤ MA_10(t-1) | -2 |
| MA_M(t) < MA_10(t) | +1 |
| MA_M(t) > MA_10(t) | -1 |
| Other Cases | 0 |

MA_M(t): Moiving Average of M values, MA_10: Moving Average of 10- values

**Oscillators Simulators**

| Conditions of CCI | Trend Forecasting Signal |
|---|---|
| CCI(t)<-150 && CCI(t)<CCI(t-1) && CCI(t-1)<CCI(t-2) && CCI(t-2)<CCI(t-3) | +2 |
| CCI(t)>150 && CCI(t)>CCI(t-1) && CCI(t-1)>CCI(t-2) && CCI(t-2)>CCI(t-3) | -2 |
| CCI(t)<-150 | +1,5 |
| CCI(t)>150 | -1,5 |
| CCI(t)<-100 | +1 |
| CCI(t)>100 | -1 |
| CCI(t)<CCI(t-1) && CCI(t-1)<CCI(t-2) && CCI(t)<0 | +0,5 |
| CCI(t)>CCI(t-1) && CCI(t-1)>CCI(t-2) && CCI(t)>0 | -0,5 |
| Other Cases | 0 |

| Conditions of Williams | Trend Forecasting Signal |
|---|---|
| WILL(t)<-99 && WILL(t)<WILL(t-1) && WILL(t-1)<WILL(t-2) && WILL(t-2)<WILL(t-3) | +2 |
| WILL(t)>-1 && WILL(t)>WILL(t-1) && WILL(t-1)>WILL(t-2) && WILL(t-2)>WILL(t-3) | -2 |
| WILL(t)<-99 | +2 |
| WILL(t)>-1 | -2 |
| WILL(t)<-98 | +1,5 |
| WILL(t)>-2 | -1,5 |
| WILL(t)<-80 | +1 |
| WILL(t)>-20 | -1 |
| WILL(t)<-80 | +0,5 |
| WILL(t)>-20 | -0,5 |
| Other Cases | 0 |

| Conditions of RSI | Trend Forecasting Signal |
|---|---|
| RSI(t)<5 && RSI(t)<RSI(t-1) && RSI(t-1)<RSI(t-2) && RSI(t-2)<RSI(t-3) | +2 |
| RSI(t)>90 && RSI(t)>RSI(t-1) && RSI(t-1)>RSI(t-2) && RSI(t-2)>RSI(t-3) | -2 |
| RSI(t)<5 | +1,5 |
| RSI(t)>90 | -1,5 |
| RSI(t)<15 | +1 |
| RSI(t)>85 | -1 |
| RSI(t)<30 | +0,5 |
| RSI(t)>70 | -0,5 |
| Other Cases | 0 |

| Conditions of Price Oscillator | Trend Forecasting Signal |
|---|---|
| PROSC(t)<-12 | +2 |
| PROSC(t)>12 | -2 |
| PROSC(t)<-9 | +1,5 |
| PROSC(t)>9 | -1,5 |
| PROSC(t)<6 | +1 |
| PROSC(t)>-6 | -1 |
| PROSC(t)<0 | +0,5 |
| PROSC(t)>0 | -0,5 |
| Other Cases | 0 |